\documentclass[useAMS,usenatbib]{mn2e}
\usepackage{amsmath} 
\usepackage{subfigure}
\usepackage{times}

\usepackage{graphicx}
\usepackage{times}%
\usepackage{natbib}
\usepackage{times}%
\title[Properties of Long Gamma-Ray Burst Host Galaxies]
      {Properties of Long Gamma-Ray Burst Host Galaxies in Cosmological Simulations}
\author[Campisi M.A. et al.]
       {M.A. Campisi$^1$\thanks{E-mail: campisi@mpa-garching.mpg.de}, 
	G. De Lucia$^1$\thanks{INAF - Astronomical Observatory of Trieste, via G.B. Tiepoli 11, I-34131 Trieste, Italy.}, L-X. Li$^1$, 
        S. Mao$^2$ and X. Kang $^3$\\
        $^1$ Max-Planck-Institut f\"ur Astrophysik, 
        Karl--Schwarzschild--Str. 1, D-85748 Garching, Germany \\ 
        $^2$ Jodrell Bank Centre for Astrophysics, Alan Turing Building, 
        University of Manchester, Manchester, M13 9PL, UK\\
	$^3$ Max-Planck-Institut für Astronomie, K\"onigstuhl 17, 69117 Heidelberg, Germany\\}

\begin{document}

\date{Accepted 2009 August 17. Received 2009 August 17; in original form 2009 January 27}
\pagerange{\pageref{firstpage}--\pageref{lastpage}} 
\pubyear{2009}

\maketitle

\label{firstpage}
\begin{abstract}
We use galaxy catalogues constructed by combining high-resolution N-body
simulations with semi-analytic models of galaxy formation to study the
properties of Long Gamma-Ray Burst (LGRB) host galaxies. We assume that LGRBs
originate from the death of massive young stars and analyse how results are
affected by different metallicity constraints on the progenitor stars.  As
expected, the host sample with no metallicity restriction on the progenitor
stars provides a perfect tracer of the cosmic star formation history. When LGRBs are
required to be generated by low-metallicity stars, they trace a decreasing
fraction of the cosmic star formation rate at lower redshift, as a consequence
of the global increase in metallicity. 
We study the properties of host galaxies up to high redshift ($\sim 9$), finding that they typically have low-metallicity ($Z<0.5 Z_{\odot}$) and that they are small ($M<10^9 M_{\odot}$), bluer and younger than the average galaxy
population, in agreement with observational data. They are also less clustered than typical $L_*$
galaxies in the Universe, and their descendents are
massive, red and reside in groups of galaxies with halo mass between $10^{13} M_\odot$ to
$10^{14} M_\odot$. 
\end{abstract}


\section{Introduction} 
\label{sec:intro}

Gamma-ray bursts (GRBs) are the most energetic explosions in the Universe
\citep{zhang04}. As such, they offer exciting possibilities to study
astrophysics in extreme conditions, e.g., radiative processes in highly
relativistic ejecta \citep[and references therein]{huang00,fan08}. Because of
their very large luminosity, GRBs represent ``cosmological'' events, which have
been detected up to $z\sim8.2$ \citep{tan09,sal09b}.
It has been proposed that some tight correlations among GRB parameters can make them  
``standard candles'' for probing the Universe 
to the high-redshift regime that supernovae Ia
cannot attain (e.g. \citealt{ghirlanda2004,dai04}, \citealt{zhang07}, and references therein. See however \citealt{li07a,li07b,li08b}).

It is well-known that the duration distribution of GRBs is
bimodal \citep{kou93}, dividing GRBs into two classes: long GRBs (hereafter
LGRBs) and short GRBs, depending on whether their durations are longer or
shorter than a few seconds.  The observed properties of host galaxies of short
and long GRBs indicate that they have different progenitors. LGRBs are
typically found in star-forming galaxies, predominantly irregular dwarf
galaxies \citep{Conselice_etal_2005, fru06, wai07}.  In contrast, short GRBs
are found in both early-type and late-type galaxies. Many models have been
proposed for explaining the origin of these two classes of GRBs. 
The currently favourite hypotheses are that short GRBs are produced by the
merger of compact objects -- between two neutron stars or between a neutron
star and a black hole \citep{li98,osh08}, while LGRBs originate from the
death of massive stars (with low metallicity), such as Wolf-Rayet stars. This
scenario for the formation of LGRBs is usually referred to as the ``collapsar
model'' \citep{Yoon_Langer_Norman_2006,yoon08, woo06b}.

Observational data are consistent with the hypothesis of a LGRB-supernova
connection: at least some LGRBs are associated with core-collapse supernovae \citep[and references therein]{gal98,li06,woo06b}. In addition,
all supernovae associated with GRBs are Type Ic, which supports the
hypothesis of Wolf-Rayet stars as progenitors of LGRBs. Because of their
connection with supernovae, LGRBs are potential tracers of the cosmic
star formation history \citep{Totani_1997,Wijers_etal_1998,mao98,Porciani_Madau_2001}.
To date, there are $~130$ GRBs with known redshift and $~50$ with estimated host galaxy stellar mass
\citep{Savaglio_etal_2008}. Given the
difficulty in detecting and localizing short GRBs, most of the observational
studies about host galaxies are for LGRBs, which will be the focus of
this work. Studies of the physical properties of GRBs are not easy as
they require deep targeted observations at high redshift. In addition, the
probability for a chance superposition of GRBs and galaxies on the sky is
significant for high-z GRBs, and $\sim 3\%$ for galaxies at $z<1.5$ \citep{campisi08}.

The observational information gathered so far indicates that most LGRBs are
found in faint star forming galaxies dominated by young stellar populations
with a sub-solar gas-phase metallicities, although there are a few host galaxies
with higher metal content \citep[and references therein]{Prochaska_etal_2004,wol07,fyn06,price07,sav03, Savaglio_2006,
Savaglio_etal_2008,sta08}.

In this work, we analyse the properties of host galaxies of LGRBs, using
a galaxy catalogue constructed by combining high-resolution
N-body simulations with a semi-analytic model of galaxy formation.
In particular, we use the models discussed in Wang et al. (2008) for two cosmological models with parameters taken from
the first-year and the third-year Wilkinson Microwave Anisotropy Probe (WMAP) \citep{spe03}
measurements. To select candidate host galaxies of LGRBs, we extract from the available semi-analytic galaxy catalogues
the information for the age and metallicity of newly formed stars and we adopt the
collapsar model. We built three
samples of host galaxies with different metallicity thresholds, and we compare the
properties of the selected galaxies with observational data, in particular the
data in \cite{Savaglio_etal_2008}. Compared with previous theoretical
studies, the simulations used in our study have the largest volume, and they also allow us
to explore the cosmological dependence. Finally, the information available from the semi-analytic catalogues
enable us to study the clustering and
descendent properties of LGRB hosts.

The paper is organised as follows. We
present in section 2 the simulated galaxy catalogues used in this work.  In
section 3, we describe the method for selecting long GRB host galaxies. We
describe our results in section 4. We discuss our results and give our conclusions in section 5.

\section{The Simulated Galaxy Catalogues}
\label{sec:sam}

\begin{table}
  \centering
  \begin{tabular}{cccccccccc} \hline \hline
    $ $ & ${\rm WMAP1}$ & ${\rm WMAP3}$ \\ \hline \hline
    $\Omega_m$ & 0.25 & 0.226 \\
    $\Omega_{\Lambda}$ & 0.75 & 0.774 \\
    $\Omega_b$ &0.045 & 0.04 \\
    $\sigma_8$& 0.9 & 0.722 \\
    $h$ & 0.73   & 0.743 \\
    $n$ & 1 & 0.947 \\
    \hline
    \hline
  \end{tabular}
  \caption{Cosmological parameters of the two simulations used in 
  \citet{Wang_etal_2008}. $\Omega_m$, $\Omega_{\Lambda}$, $\Omega_b$ represent 
  the density of matter, dark energy, and baryons respectively. $\sigma_8$ and 
  $n$ are the amplitude of the mass density fluctuations, and the slope of the
  initial power spectrum. The Hubble constant is parameterised as 
  $H_0 = 100\, h\, {\rm km\, s^{-1} Mpc^{-1}}$.}  
\label{tab:cosmparam}
\end{table}

In this study, we use the galaxy catalogues constructed by 
\citet{Wang_etal_2008} for two simulations with cosmological parameters from 
the first and third-year WMAP results. The two sets of cosmological parameters
are listed in Table~\ref{tab:cosmparam}. As discussed
in \citet{Wang_etal_2008}, the most significant differences between WMAP1 and
WMAP3 cosmological parameters are a lower value of $\sigma_8$ and a redder
(smaller) primordial power spectrum index $n$ in WMAP3, resulting in a
significant delay for structure formation. Both simulations correspond to a box
of $125\,h^{-1} {\rm Mpc}$ comoving length and a particle mass $8.6\times 10^8$
(WMAP1) and $7.8\times 10^8\,{\rm M}_{\odot}$ (WMAP3). The softening length is
5 $h^{-1}$ kpc in both simulations (see Table~2 in Wang et
al. 2008). Simulation data were stored in 64 outputs, that are
approximately logarithmically spaced in time between $z=20$ and $z=1$, and
linearly spaced in time for $z<1$. Each simulation output was analysed with
the post-processing software originally developed for the Millennium
Simulation \citep{Springel_etal_2005}.

Merging history trees for self-bound structures extracted from the simulations
were used as input for the Munich semi-analytic model of galaxy
formation. Interested readers are referred to \citet{Croton_etal_2006},
\citet{DeLucia_Blaizot_2007} and references therein for details on the
physical processes explicitly modelled. Previous work has shown that the galaxy
population predicted by this particular model provides a reasonably good match
with the observed local galaxies properties and relations among stellar mass,
gas mass, and metallicity \citep*{DeLucia_Kauffmann_White_2004}, luminosity,
colour, morphology distributions \citep{Croton_etal_2006,DeLucia_etal_2006},
and the observed two-point correlation functions
\citep{Springel_etal_2005,Wang_etal_2008}. In addition, \citet{Kitzbichler_White_2007} have
shown that the model also agrees reasonably well with the observed galaxy
luminosity and mass function at higher redshift. 

We remind the reader that the models discussed in \citet{Wang_etal_2008} adopt
the same physics, but different combinations of model parameters are used for
the simulations with WMAP1 and WMAP3 cosmology. The WMAP1 simulation uses the
same parameters (and physical model) adopted in
\citet{DeLucia_Blaizot_2007}. For the WMAP3 simulation, the model adopts lower 
supernovae and AGN feedback efficiencies in order to compensate for the delay
in structure formation obtained with a lower $\sigma_8$. This combination of
model parameters corresponds to the WMAP3B model used
in \citet{Wang_etal_2008}. The alternative model WMAP3C used in that paper
leads to very similar results and so will not be discussed further in this
paper. In the following, we limit our analysis to galaxies with stellar mass
larger than $2\times 10^8\,{\rm M}_{\odot}$, which is above the resolution
limit of the N-body simulations used.

We note that the most recent cosmological model from the five-year data
of WMAP is between WMAP1 and WMAP3, so the results from the two simulations
used here are expected to bracket results from a simulation with the 5-year WMAP
cosmology.

\section{Identification of LGRB host galaxies}
\label{sec:method}

In order to identify candidate GRB host galaxies, we adopt the collapsar model
for LGRBs: all young stars with mass $>30\,M_{\odot}$ ending their life with a
supernova should be able to create a BH remnant. If the collapsar has high
angular momentum, the formation of the BH is accompanied by a GRB
event \citep{Yoon_Langer_Norman_2006,yoon08}. As mentioned in
Sec.~\ref{sec:intro}, recent studies on the final evolutionary stages of
massive stars have suggested that a Wolf-Rayet (WR) star can produce a LGRB if its
mass loss rate is small, which is possible only if the metallicity of the star
is very low. When metallicities are lower than $\sim 0.1-0.3\,Z_{\odot}$, the
specific angular momentum of the progenitor allows the loss of the hydrogen
envelope while preserving the helium
core \citep*{woo06b,Fryer_Woosley_Hartmann_1999}. The loss of the envelope reduces
the material that the jet needs to cross in order to escape, while the helium core
should be massive enough to collapse and power a GRB.

In order to construct our host galaxy sample, we have extracted from the
available semi-analytic catalogues the information about the age and
metallicity of all stars. We have then created the following host samples:
\begin{enumerate}
\item HOST1, obtained by selecting galaxies containing stars with
  age $< t_c = 5\times 10^7 {\rm yr}$;
\item HOST2, including galaxies with stars of age $< t_c$ and
  metallicity $Z\leq0.3Z_{\odot}$;
\item HOST3, defined by selecting galaxies containing stars with age $< t_c$ 
  and metallicity $Z\leq0.1Z_{\odot}$.
\end{enumerate}

In order to count the number of GRB events in each galaxy, 
we make use of two important pieces of information: (a) the 
rate of GRB with respect to the SNe explosions (without any cut in
metallicity for the progenitors stars); (b) the rate of very massive
stars (producing remnant BHs) as a function of redshift and metallicity with respect to the total number of SNe events. In this way we are able to count how many BHs with low metallicity progenitors will produce GRBs.\\

We assume a Salpeter
Initial Mass Function (IMF) and compute the number of stars ending their
lives as supernovae (SNe) or as black-holes (BHs) per unit mass of stars
formed by:
\begin{equation}
  N = \dfrac {	\displaystyle \int_{\small m_{\rm *, min}}^{\small m_{\rm *, max}} \phi(m_*)\,dm_*}{\displaystyle \int_{0.1}^{\small m_{\rm *, max}} m_* \phi(m_*)\,dm_*},
\label{eq:eq1}
\end{equation}
where $\phi(m_*)$ is the IMF, $m_{\rm *, min}$ is the minimum initial mass to
form a supernova or a black hole, and $m_{\rm *, max}$ is the upper limit of
the mass function. We take $m_{\rm *, min}$ to be $8\,M_{\odot}$ for SNe and
$30\,M_{\odot}$ for BHs \citep{Fryer_Woosley_Hartmann_1999}, and $m_{\rm *,
max} = 100\,M_{\odot}$ (the lower limit of the IMF is taken to be $0.1
M_\odot$). Eq.~\ref{eq:eq1}, evaluated for progenitor stars of all
metallicities, provides the numbers listed in Table~\ref{tab:tab2}. 
\begin{table}
  \begin{center}
  \begin{tabular}{|cccc|} 
  \hline
  \hline
  &$ N [M_{\odot}^{-1}]$ & $m_{*, {\rm min}}$ & $m_{*, {\rm max}}$ \\ \hline
  SNe&$7.421\times10^{-3}$&8&100\\
  BH&$1.035\times10^{-3}$&30&100\\
  \hline  
  \hline  
  \end{tabular}
  \end{center}
  \caption{Number of SNe and BHs per solar mass of stars formed, using a
    Salpeter IMF.}  
\label{tab:tab2}
\end{table}
\begin{figure}
  \centering
  \includegraphics[scale=0.4, angle=-90]{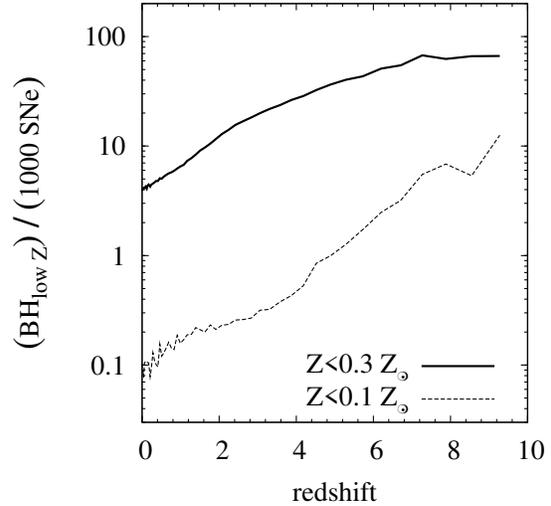}
  \caption{Number of low-Z BHs per 1000 SNe as a function of redshift, for
    progenitor stars with metallicity lower than $0.3\,Z_{\odot}$ and
    $0.1\,Z_{\odot}$. Results are shown for the model WMAP3B. Similar results
    are obtained in the WMAP1 case.}
\label{fig:rate}
\end{figure}
Without any restriction on the metallicity of the progenitor stars, the
relative number of BHs and supernovae is $\sim 14$ per cent (140 BHs per 1000
SNe).
When considering the metallicity threshold for BHs, the rate is reduced and
varies with redshift \citep{lan06,wol07}. This is illustrated in
Fig.~\ref{fig:rate} for the two different metallicity thresholds adopted in our
study. At low redshift, there are from 0.1 to 6 low-metallicity BHs formed per
1000 SNe. The rate grows with increasing redshift, reaching values between
$\sim 10$ ($Z < 0.1\,Z_{\odot}$) and $\sim 70$ ($Z < 0.3\,Z_{\odot}$) at
redshift $\sim 9$.
 
Integrating over the redshift, Fig.~\ref{fig:rate} allows us to
calculate the average ratio between the number of black holes from different
progenitors and the number
of supernovae in the Universe. However, not every black hole will produce a LGRB. To normalise the LGRB abundance, we 
assume that the rate of GRB per SNe is on average (over all cosmic times)
of about 1 GRB event every 1000 SNe
\citep{Porciani_Madau_2001,lan06}. The
number of LGRBs relative to the black holes in the three samples (with
or without the metallicity cutoff) is given by
\begin{equation}
R_{\rm GRB} = \frac{{\rm GRBs}}{{\rm BHs_{low Z}}} = \frac{{\rm GRBs}}{1000 {\rm SNe}} \times 
             \langle \frac{1000 {\rm SNe}}{{\rm BHs_{low Z}}} \rangle.
\end{equation}
where the last term is obtaining by integrating over redshift (or
cosmic time). For
the HOST1 sample, the last term is a constant (1/140), and thus $R_{GRB}\sim 0.007$;
for the other two samples, we obtain $R_{\rm GRB} \sim 0.056$ and $\sim 1$.

For each galaxy in the simulation box, we can count how many BHs are
produced from low metallicity progenitors and then obtain the
corresponding number of LGRBs:
\begin{displaymath}
  N_{{\rm LGRBs}} = {\rm BHs_{low Z}}\times R_{\rm {GRB}}.
\end{displaymath}

It is important to note that our model provides only an upper limit to the
number of LGRB events because the formation of BHs and low metallicity are only
two of many requirements for the production of LGRBs. E.g., high spin of
the progenitor star is another requirement that cannot be handled in our model.
In the following, we consider each galaxy hosting at least one GRB event
($N_{{\rm LGRBs}}\geq 1$) as a `host galaxy'. \\
There are relatively few galaxies with $N_{{\rm LGRBs}}< 1$ and their inclusion does not
change our results significantly.\\
We stress that we are not using the (average) galaxy metallicity to select
our host galaxy samples, but the metallicity of each `pocket' of stars formed
at each time-step. Stars are assumed to form with the metallicity of the
interstellar medium at the time of star formation, and the model adopts an
instantaneous recycling approximation for metal enrichment. So, the gas-phase
metallicity of host galaxies will generally be higher than the metallicity
threshold we have adopted for our samples.\\
Note that the LGRB rate computed above is not directly comparable to the
observed rate because that would require us to take into account many unknown
factors like the jet angle, and to include any possible observational bias 
\citep[see][]{Lapi_etal_2008,li08a}.

\section{Results}

In this section, we discuss the physical properties of LGRB hosts selected
using the procedure described in Sec.~\ref{sec:method}. In particular, we
compare the LGRB rate to the cosmic star formation rate (Sec.~\ref{sec:sfr}),
and the properties of LGRBs hosts to the global properties of galaxies at the
same cosmic epoch (Sec.~\ref{sec:compare}). When we study the average or median property of the host galaxies we 
weigh each host by the
likelihood that it contains a GRB. That 
 allow us to compare with observed data
since the galaxies which host many GRBs are
proportionally more likely to appear in any given observed sample.
 We also study the typical
environment (Sec.~\ref{sec:environment}) and evolutionary stage
(Sec.~\ref{sec:evolve}) of LGRB host galaxies.

\subsection{The cosmic star formation rate versus the LGRB rate}
\label{sec:sfr}

\begin{figure*}
  \centering
  \includegraphics[scale=0.38,angle=-90]{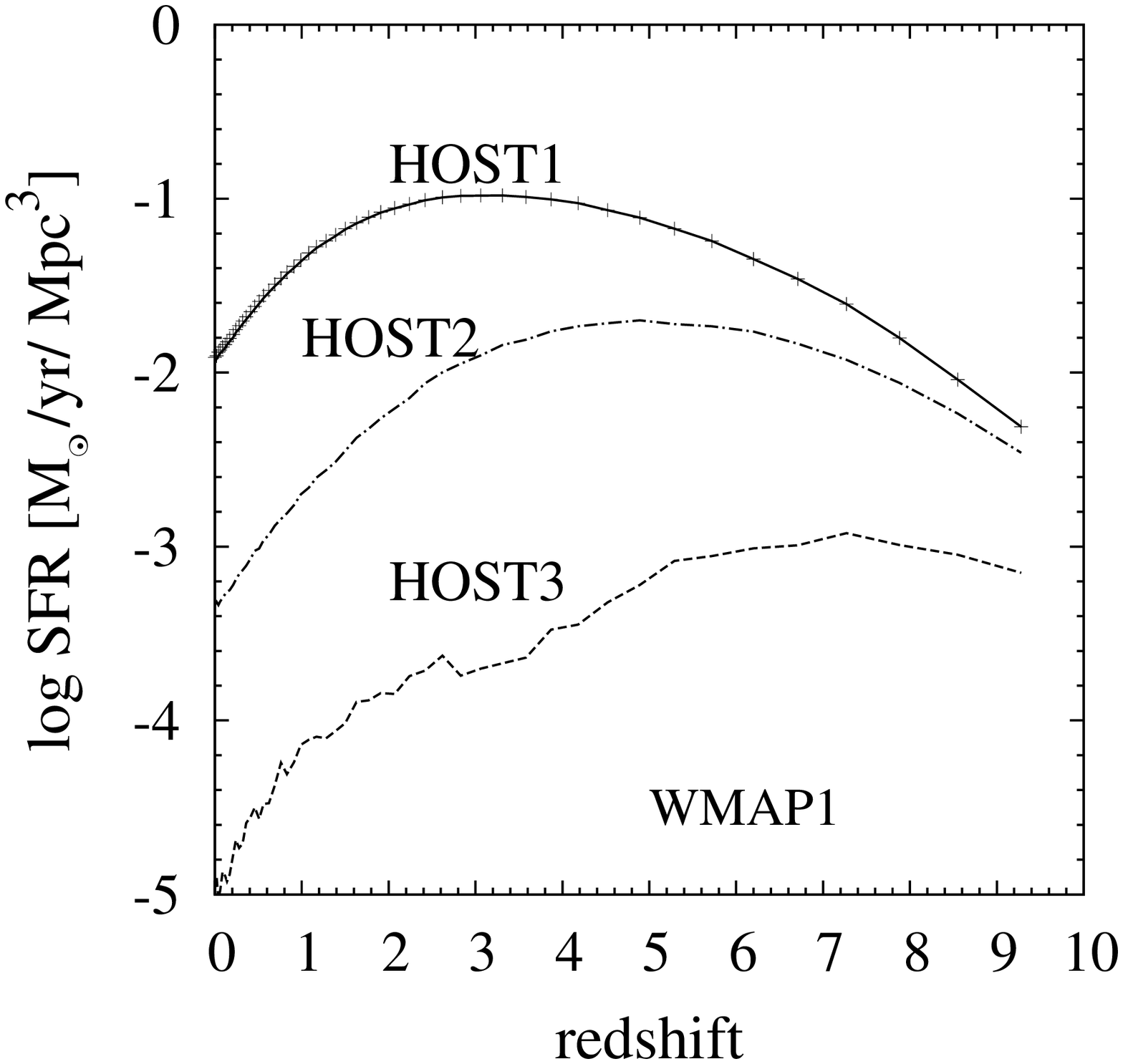}
  \includegraphics[scale=0.38,angle=-90]{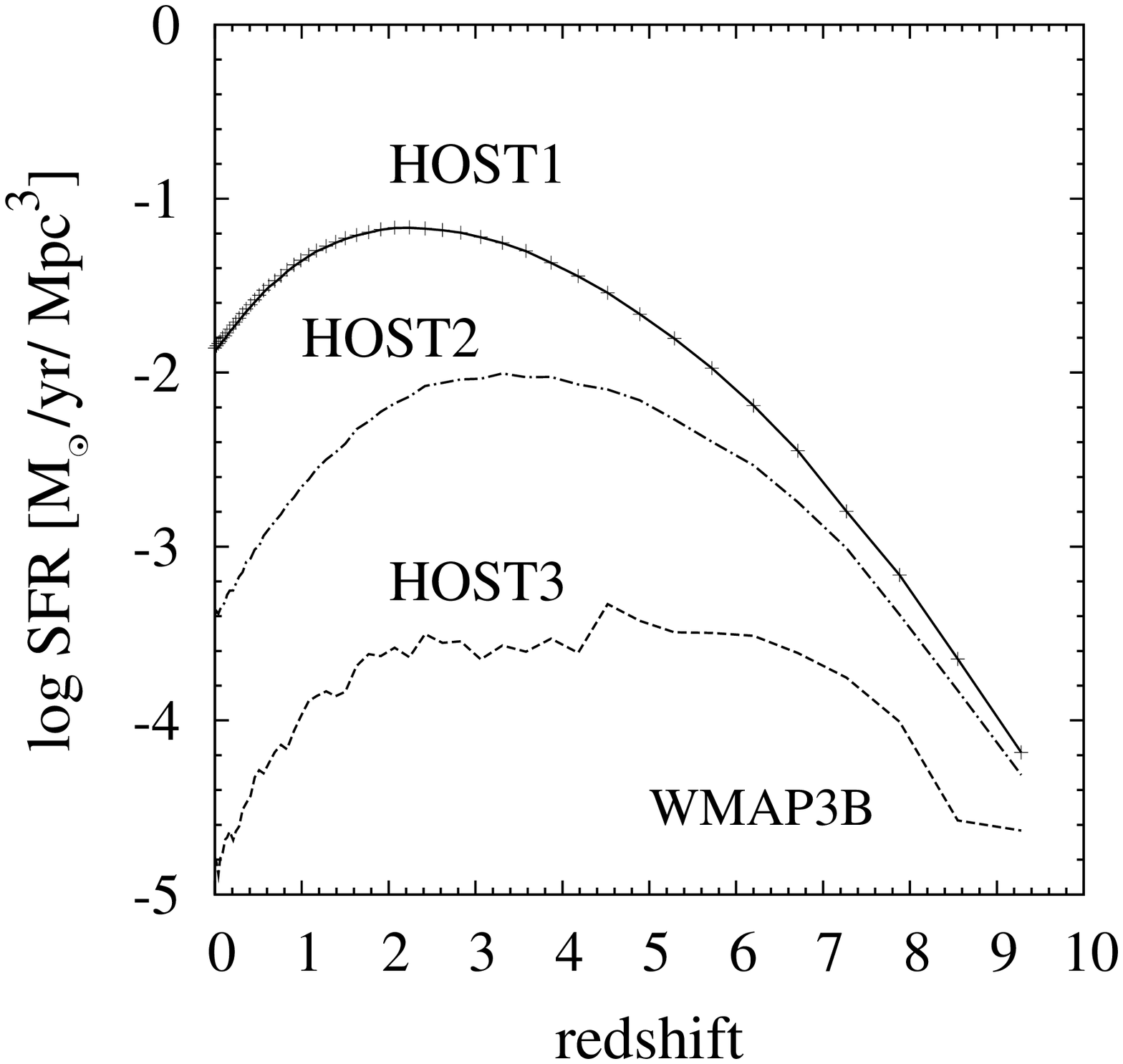}
  \caption{${\rm Log}\,SFR\,[M_{\odot}\,{\rm yr}^{-1}\,{\rm Mpc}^{-3}]$
  as a function of redshift, computed for all galaxies in the simulation box (solid
  line). Left and right panels correspond to the WMAP1 and WMAP3 simulations 
  respectively. The dot-dashed and dashed lines corresponds to the HOST2 and
  HOST3 samples. The sample with no threshold on metallicity (HOST1) traces
  exactly the global star formation rate measured considering all galaxies in
  the simulated boxes (solid line).}
\label{fig:sfr}
\end{figure*}

The collapsar model links LGRBs to the evolution of single massive stars whose
lifetimes are negligible on cosmological scales. If no other condition is
required for producing a LGRB event, then the rate of LGRBs should be an
unbiased tracer of the global star formation in the Universe \citep[e.g.][and
references therein]{Totani_1997,Wijers_etal_1998,mao98,Porciani_Madau_2001,bro02,fyn06,pri06,Savaglio_2006,tot06,pro07,li08a}.
\cite{fyn08} have recently suggested that GRB and Damped Lyman-Alpha samples, in contrast with magnitude limited samples, provide an almost complete census of $z\sim3$ star-forming galaxies. We note, however, that the sample used in this study are biased against high-metallicity and dusty systems.\\
However, both observations and theoretical studies indicate that the
metallicity of the progenitor star plays an important role in setting the
necessary conditions for a LGRB explosion. 
In this case, the rate of LGRBs is
expected to be a biased tracer of the cosmic star formation rate. This is
demonstrated explicitly in Fig.~\ref{fig:sfr}, which compares the cosmic star
formation rate obtained using all galaxies in the simulation box to that
obtained for the three LGRB host samples defined in the Sec.~\ref{sec:method}.

The sample with no threshold on metallicity (HOST1) traces exactly the global
star formation rate (solid line in Fig.~\ref{fig:sfr}). This is not the case
for the two samples with metallicity thresholds (HOST2 - dot-dashed line and
HOST3 - dashed line). For the HOST2 and HOST3 samples, the LGRB rate peaks at
higher redshift than the cosmic star formation rate, as a consequence of the
global decrease of metallicity with increasing redshift. At higher redshift,
the mean metallicity of the intergalactic medium is lower, which implies that a
larger fraction of stars form below the metallicity thresholds adopted for HOST2
and HOST3. Therefore the deviation of the LGRB rate from the star formation
rate decreases with increasing redshift. At $z\sim 9$, the HOST2 sample
measures about 99 per cent of the global star formation density. The fraction
decreases to about 30 per cent at $z\sim 5$ and to only about 10 per cent at
present. For the HOST3 sample, the bias is even stronger because of the lower
metallicity threshold: it traces only 30 per cent of the global star formation
density at $z\sim 9$, and less than 10 per cent at $z<5$. The results shown in
Fig.~\ref{fig:sfr} are in qualitative agreement with recent observational
estimates \citep{Kistler_etal_2008}, and with recent theoretical studies also
based on the collapsar model (\citealt*{Yoon_Langer_Norman_2006},
\citealt{Cen_Fang_2007}, \citealt{Nuza_etal_2007}, \citealt{Lapi_etal_2008}).

Finally, Fig.~\ref{fig:sfr} also shows that adopting a cosmological model
compatible with third-year WMAP measurements (right panel), a delay is produced
in the cosmic star formation rate, due to the delay in structure
formation. Except for this delay, the predicted trends are
the same for the WMAP1 and WMAP3 simulations. In the following, we will only
show results obtained using the WMAP3 simulation as those obtained for the
WMAP1 simulation are very similar. In addition, we will focus only on the two
host samples with metallicity thresholds (HOST2 and HOST3).

\begin{figure*}
  \centering
\includegraphics[scale=0.36,angle=-90]{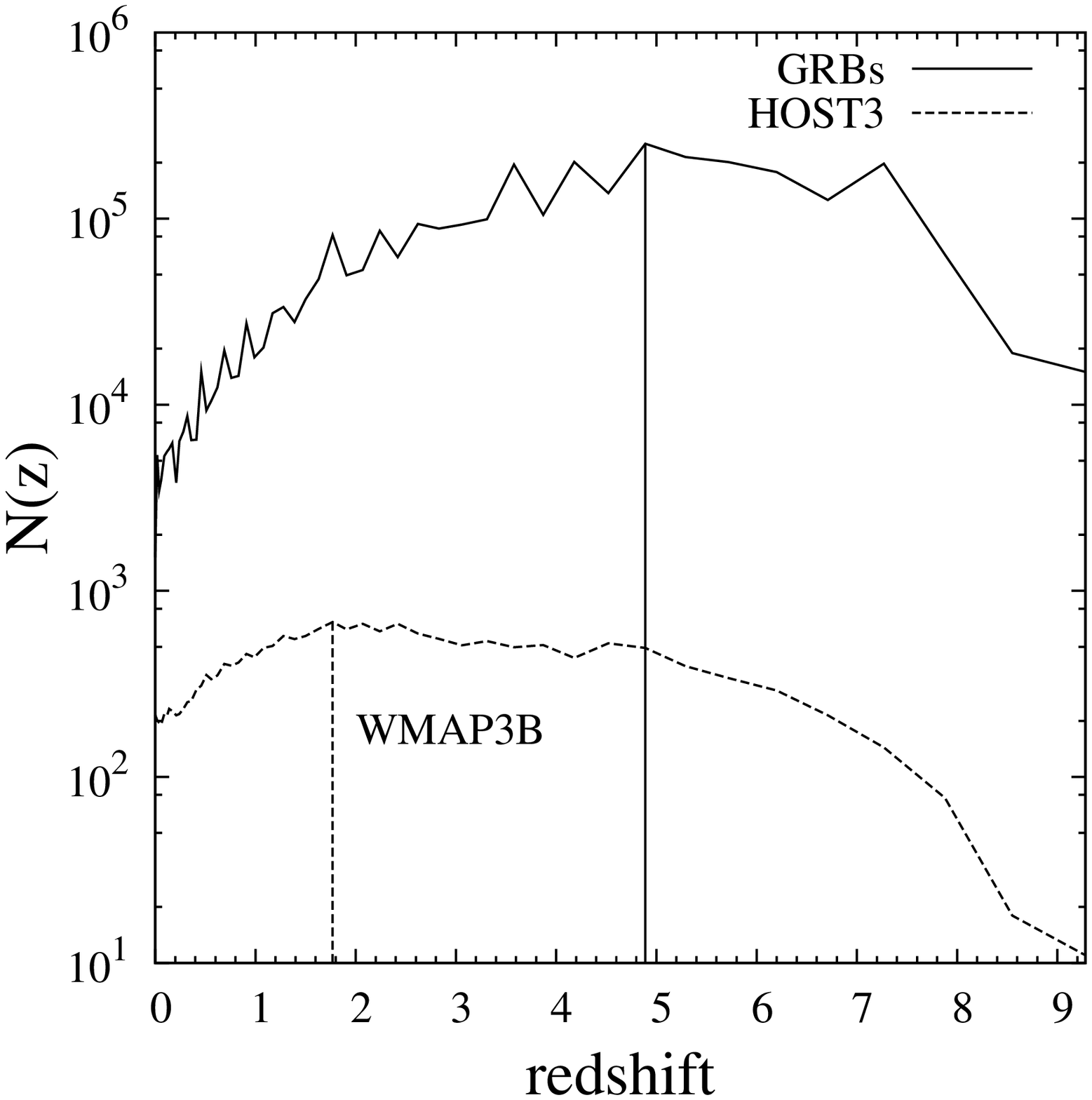} 
\includegraphics[scale=0.38,angle=-90]{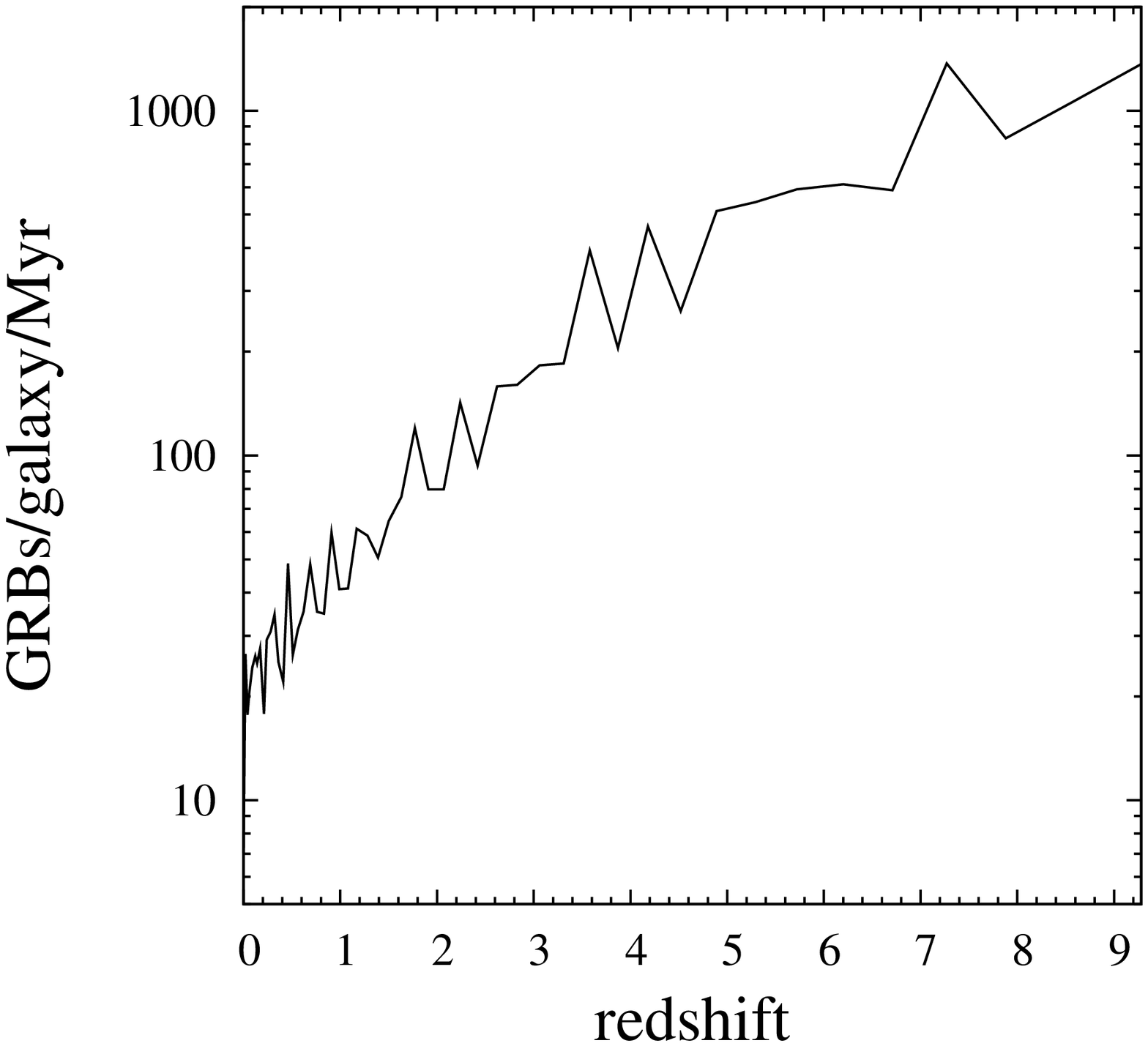} 
  \caption{Left panel: redshift distribution of LGRB events (solid line) and of
  host galaxies (dashed line), for the sample HOST3 (with metallicity
  threshold $0.1\,Z_{\odot}$). The vertical lines indicate the peaks of the
  distributions. Right panel: the rate of LGRBs per galaxy per Myr. Both
  panels correspond to results from the WMAP3 simulation.} 
\label{fig:loggrb_myr}
\end{figure*} 

We remind the reader that we consider as `hosts' all galaxies which can host at
least one LGRB event between two simulation output. Since galaxies at higher redshift have lower
metallicities and form stars at higher rates, the rate of LGRBs per host galaxy
increases rapidly with redshift. The left panel in Fig.~\ref{fig:loggrb_myr}
shows the redshift evolution of the number of LGRBs (solid line) and of host
galaxies (HOST3 - dashed line). The two vertical lines indicate the peaks of
the distributions: the number of LGRBs peaks at $z\sim 5$, while the
number of host galaxies is maximum at $z\sim 2$. The right panel of
Fig.~\ref{fig:loggrb_myr} shows the rate of LGRBs per galaxy and per Myr
computed for the WMAP3 simulation and for the HOST3 sample. The predicted rate
of LGRBs decreases from $\sim 1000$ at $z \sim 9$ to about $10\,{\rm
Myr}^{-1}\,{\rm galaxy}^{-1}$ at $z \sim 0$, in agreement with calculations by
\citet{Fryer_Woosley_Hartmann_1999}.
Note that the LGRB rate computed above is not directly comparable to the
observed rate because that would require us to take into account many unknown
factors like the jet angle, and to include any possible observational bias 
\citep[see][]{Lapi_etal_2008,li08a}.

\subsection{Physical properties of LGRB host galaxies}
\label{sec:compare}

A number of recent papers have studied the physical properties of LGRB host
galaxies using deep observations covering a large wavelength range, both in
imaging and in spectroscopy. These studies have revealed that LGRB host
galaxies are typically faint and star forming galaxies, dominated by young and
metal-poor stellar populations \citep{lef03,fru06,wai07,Savaglio_etal_2008}. In this section, we analyse the physical
properties of our model host galaxies, and compare these with the
properties of the average galaxy population, and with observational estimates.

\begin{figure*}
  \centering
  {\includegraphics[scale=0.55, angle=-90]{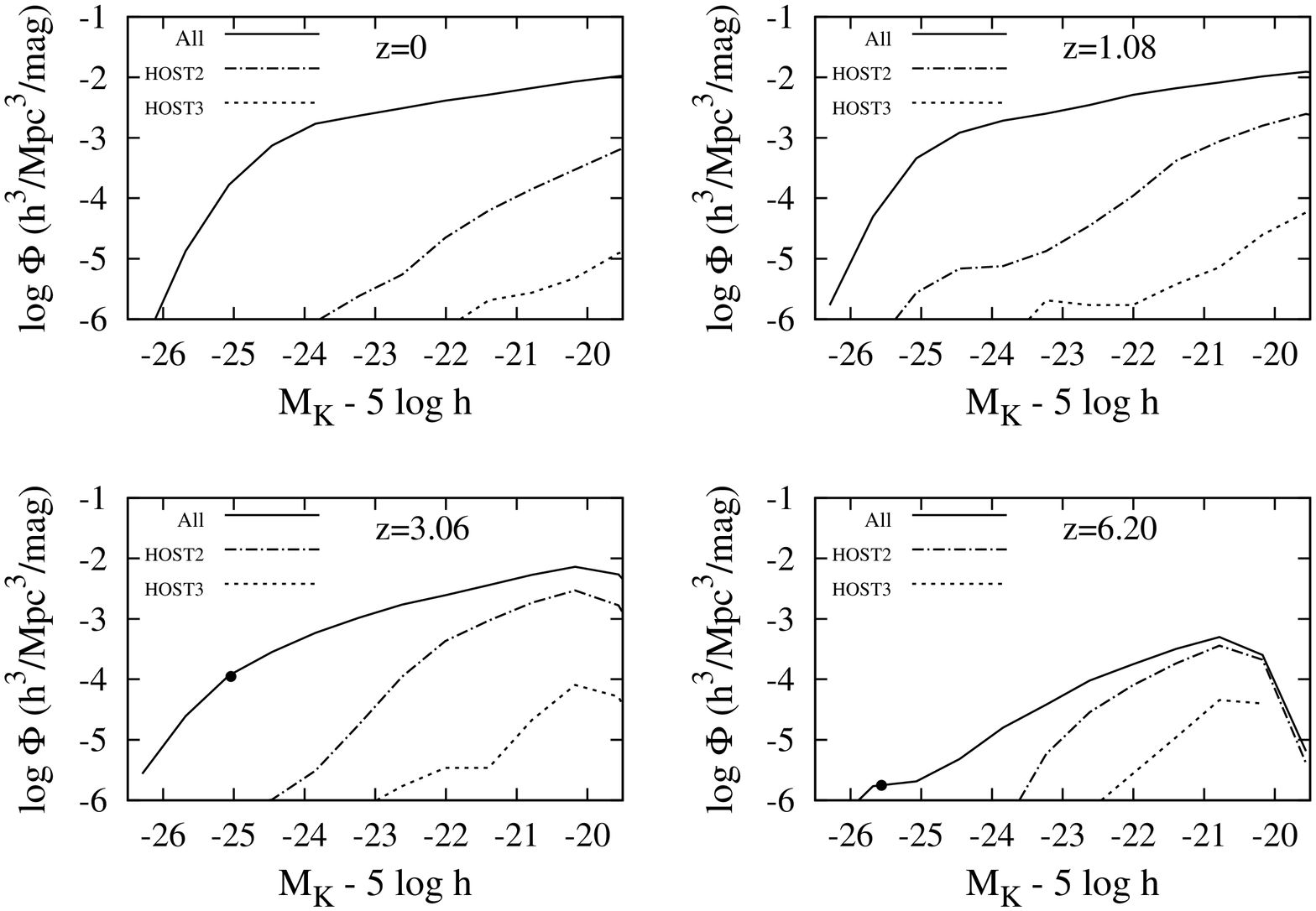}}
  \caption{Luminosity function of the HOST2 (dot-dashed line) and the HOST3
  (dashed line) samples, compared to the galaxy luminosity function measured
  using all galaxies in the simulation box (solid line). Different panels are
  for different redshifts: 0 (top left panel), 1.08 (top right panel), 3.06
  (bottom left panel), and 6.20 (bottom right panel). In the last two panel the dark bullet shows the  characteristic luminosity $L_*$.}
  \label{fig:lumfunc}
\end{figure*} 

Fig.~\ref{fig:lumfunc} shows the K-band rest-frame luminosity function of host
galaxies for the HOST2 (dot-dashed line) and the HOST3 (dashed line) samples,
compared with the galaxy luminosity function measured considering all galaxies
in the simulation box, at different redshift. At all redshifts, LGRB host
galaxies have luminosities well below the characteristic luminosity $L_*$, in
agreement with observational measurements. While the total luminosity function
evolves strongly with the redshift (particularly beyond $z\sim 1$), the number
densities and the range of luminosities of LGRB host galaxies vary more mildly,
due to the fact that at higher redshift a larger fraction of the whole galaxy
population can host LGRBs.

Recent observational studies have focused on the stellar mass distribution of 
GRB host galaxies. \citet{CastroCeron_etal_2008} have found that the typical
stellar mass of host galaxies is smaller than the stellar mass of field
galaxies at the same redshift.  For a sample of $30$ LGRB hosts, they provide
estimates of the stellar mass between $10^7$ and $10^{11} \, M_{\odot}$, with a
mean value of $M_*\sim10^{9.7}\,M_{\odot}$. About $70$ per cent of the host
galaxies in their sample have stellar mass $M_*<10^{10.1}\, M_{\odot}$.
Similar results have been found by \citet{Savaglio_etal_2008}. Using a sample
of 46 GRB hosts -the largest sample so far- they estimate a median
stellar mass of $10^{9.3}\,M_{\odot}$, and find that about $83$ per cent of the
studied systems have stellar mass between $10^{8.5}$ and $10^{10.3}\,M_{\odot}$.

\begin{figure}
  \centering
  \includegraphics[scale=0.45,angle=-90]{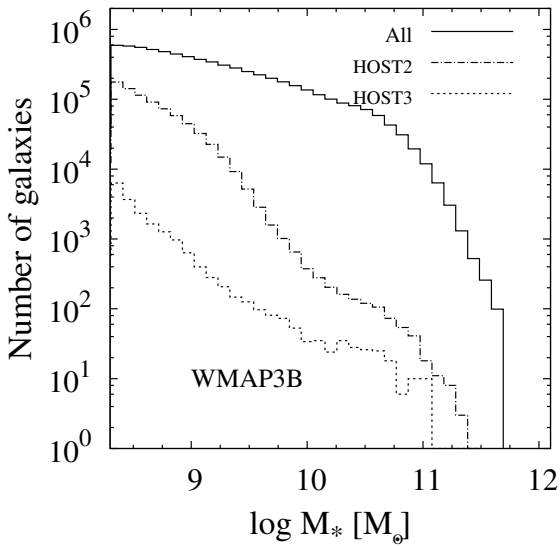}
  \caption{Galaxy mass function for the HOST2 (dot-dashed line) and HOST3
  (dashed line) samples, compared to the galaxy mass function measured using
  all galaxies in the simulation box (solid line). All galaxies with $z\leq 9$ have been used in this figure.}
  \label{fig:massfunc}
\end{figure}

In Fig.~\ref{fig:massfunc} we compare the galaxy mass distribution for model
host galaxies from the HOST2 and HOST3 samples to the distribution obtained considering
all galaxies in the simulated box. For this figure, all galaxies and hosts at
all redshifts up to $z\sim 9$ have been used. Fig.~\ref{fig:massfunc} shows that LGRB host galaxies have typically
low mass with a small fraction of them having stellar mass up to $\sim
10^{11}\,M_{\odot}$. About 90 per cent of the host
galaxies have stellar mass $<10^{9}\,M_{\odot}$ and  $<10^{10}\,M_{\odot}$ for HOST3 and HOST2 respectively. 

\begin{figure}
  \centering
   {\includegraphics[scale=0.45,angle=-90]{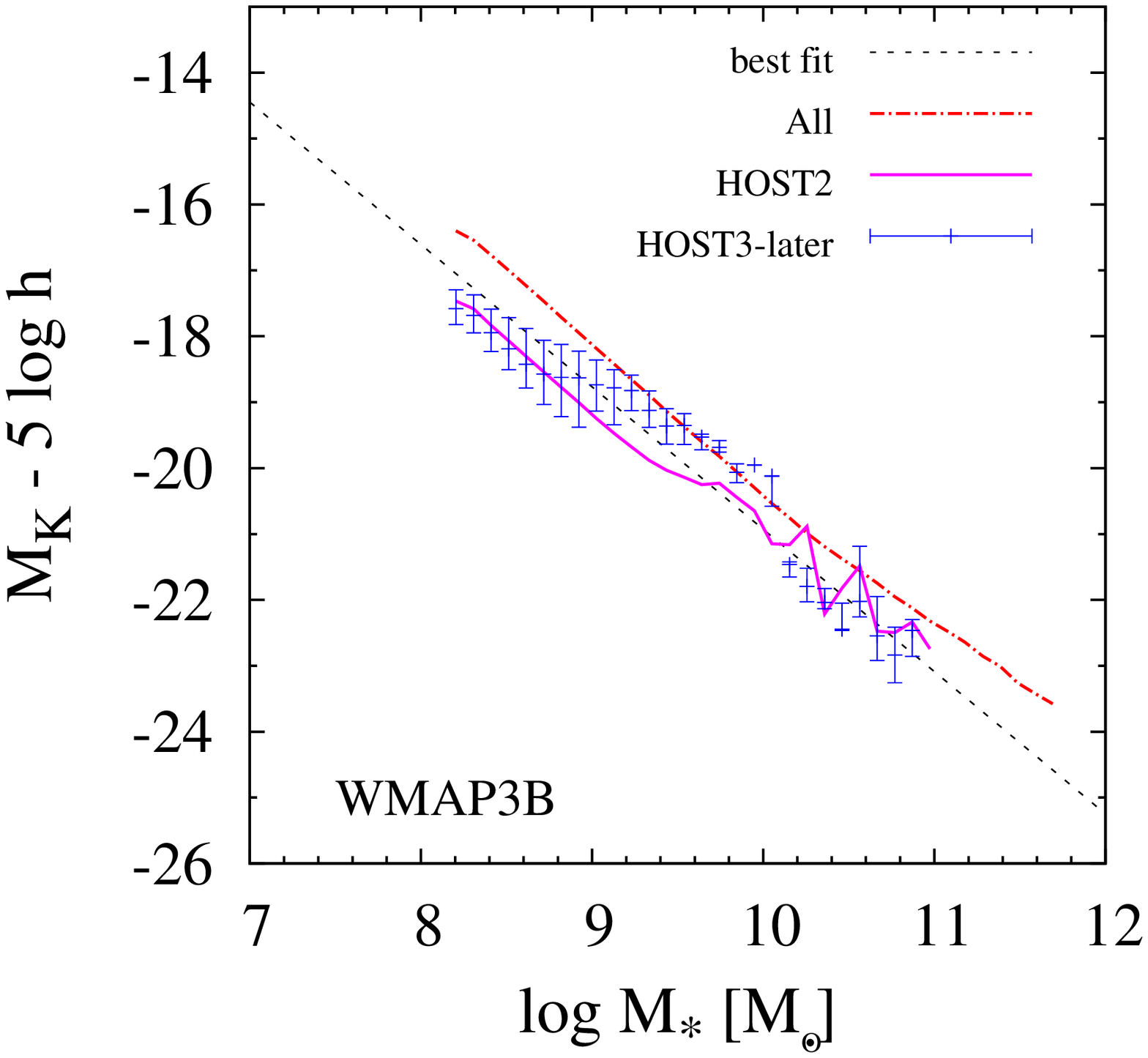}}
  \caption{Mean K-band absolute magnitude as a function of stellar mass for
    all galaxies in the simulation (red line), for galaxies in the
    HOST2 sample (green), and for galaxies in the HOST3 sample (blue). For the HOST2 and HOST3 samples we compute the average of $M_K$ weighing each host with the total number of LGRBs. Black
    symbols correspond to observational data from \citet{Savaglio_etal_2008},
    and the dotted black line is the fit provided by the same author ($
      \log\,M_* = -0.463\times M_K - 0.102$).}
\label{fig:mass_colour}
\end{figure}

It is well known that the galaxy stellar mass is tightly correlated with the
rest-frame K-band luminosity. \citet{Savaglio_etal_2008} have shown that this
relation applies to GRB host galaxies as well, but they argue that GRB galaxies
have on average higher luminosity than ``field'' galaxies with the same stellar
mass, implying a lower $M_*/L_{\rm K}$ ratio, as expected for younger galaxies.
We compare results from our model to observational measurements in
Fig.~\ref{fig:mass_colour}. The dashed black line is the best-fit to the
observational data by \citet{Savaglio_etal_2008}: ${\rm log}\,M_* =
-0.463\times M_K-0.102$. The red line in Fig.~\ref{fig:mass_colour} shows the
mean luminosity-mass relation obtained by using all galaxies in the simulation
boxes up to $z\sim 9$. The pink line shows
the mean value for host galaxies in the HOST2 sample, and the blue line
corresponds to the mean value obtained for the HOST3 sample. To compute the average of $M_K$ we 
weigh each host by the likelihood that it contains a GRB. For this sample, we also show the quartiles of the distribution. We note
that \citet{Savaglio_etal_2008} adopt a Baldry \& Glazebrook IMF for their
stellar mass estimates, while the model used in this study adopts a Chabrier
IMF to compute model magnitudes.  In order to compare model results with observational estimates, we have
decreased the observed stellar mass by a factor $1.3$.
Fig.~\ref{fig:mass_colour} shows that the K-band absolute magnitude distribution
of simulated GRB host galaxies is in good agreement with observations. It also
shows that, on average, host galaxies have stellar masses which are lower,
although with a large scatter, than ``typical'' galaxies with the same mass, in
agreement with observational findings.

\begin{figure}
  \centering
   {\includegraphics[scale=0.4,angle=-90]{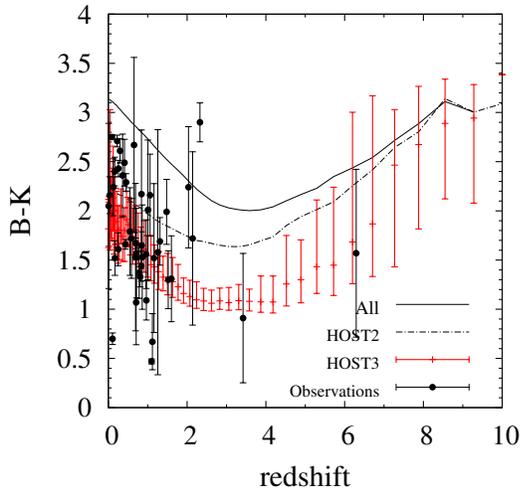}}
  \caption{Median B-K colour as a function of redshift for observed GRB hosts
  (black crosses), all galaxies in the simulation (solid line), galaxies from 
  the HOST2 sample (dot-dashed line), and galaxies from the HOST3 sample (red 
  crosses). For the HOST2 and HOST3 samples we compute the median weighing each host with the total number of LGRBs.
}
  \label{fig:colour}
\end{figure}

In Fig.~\ref{fig:colour}, we compare the median colour of model LGRB host galaxies
with observational measurements by \citet[][shown as black
symbols]{Savaglio_etal_2008}. Model results indicate that GRB galaxies are
typically bluer than the average galaxy population at the same redshift. We
note that the observed colours exhibit a quite large scatter, probably due to
the unknown dust extinction. 

\begin{figure}
  \centering
   \includegraphics[scale=0.45,angle=-90]{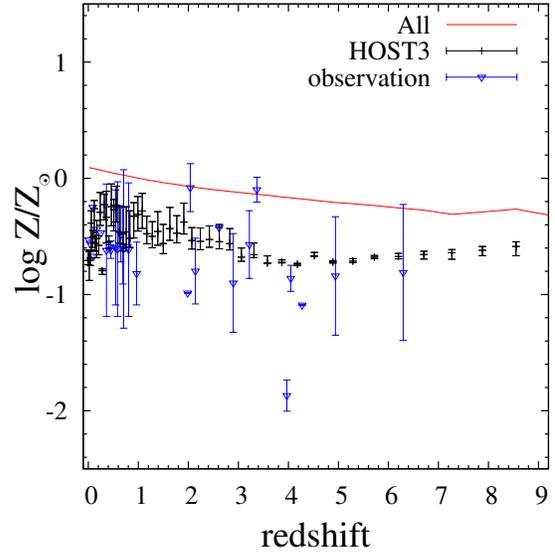}
  \caption{Median gas metallicity as a function of redshift for all galaxies in the simulation box
  (red line), and for galaxies from the HOST3 sample (dark crosses, error bars
  show the quartiles of the distributions). For HOST3 sample we compute the median gas metallicity weighing each host with the total number of LGRBs. The blue triangles are
  observational measurements for GRB-DLAs host galaxies
  from \citet{Savaglio_2006} and for the GRB host galaxies studied
  in \citet{Savaglio_etal_2008}.
}
  \label{fig:metalz}
\end{figure} 

Fig.~\ref{fig:metalz} shows the median gas
metallicity evolution for the HOST3 sample (black crosses) compared with the
observational estimates for the GRB-DLAs studied in Savaglio et al. (2006, 2008
- blue triangles with error bars a few show the lower and upper-branch
metallicity solution in Savaglio et al. 2008). In order to compare with observations we weigh each host with the total number of LGRBs. The red line in Fig.~\ref{fig:metalz} shows
the median metallicity obtained using all galaxies in the simulation box. The
figure shows that the metallicity of model galaxies (both ``normal'' and hosts)
does not evolve significantly with redshift. The observational measurements
exhibit a large scatter and have typically large uncertainties. Within these,
model predictions are in relatively good agreement with observational data. It
should be noted that the lack of evolution in the gas-phase metallicity of the
HOST3 sample is essentially due to our selection method. In order to enter this
sample, host galaxies must have young stars with metallicity lower than
$Z\leq0.1Z_{\odot}$. In the model used in this study, stars form with the
metallicity of the gas-phase component, so the adopted selection requires host
galaxies to have gas-phase metallicity close to the adopted threshold (it will
be typically higher because the model adopts an instantaneous recycling
approximation).

\begin{figure}
  \centering
  {\includegraphics[scale=0.45,angle=-90]{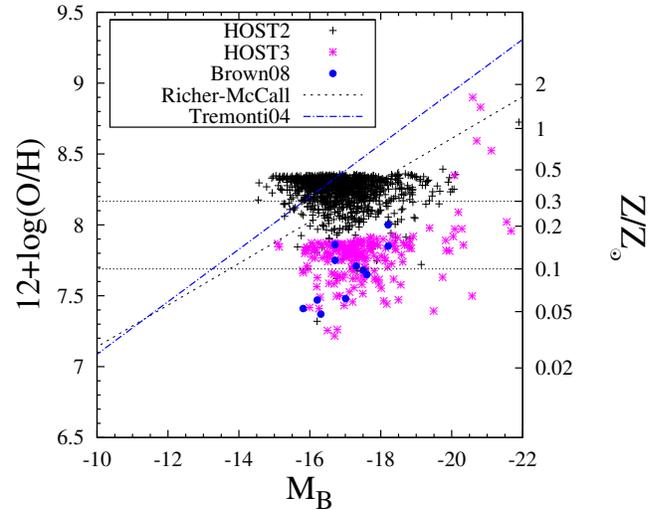}}
  \caption{Gas metallicity as function of the B-band luminosity for model
  galaxies with $z<0.1$ from the HOST3 (magenta asterisks) and HOST2 (black
  crosses) samples, compared with observational measurements by \citet[][blue
  circles]{Brown_Kewley_Geller_2008} for a sample of metal-poor galaxies. 
  The blue-dashed and black-dotted lines show the metallicity-luminosity
  relation measured by \citet{Richer_McCall_1995} for a sample of dwarf
  galaxies, and by \citet{Tremonti_etal_2004} for SDSS star forming galaxies.}
  \label{fig:metallum}
\end{figure}

Another important and well studied relation is the luminosity-metallicity
relation. This has been measured for a very large sample of star forming
galaxies from the Sloan Digital Sky Survey (SDSS)
by \citet{Tremonti_etal_2004}, and for smaller samples of galaxies by other
authors. In particular, \citet*{Brown_Kewley_Geller_2008} have recently
suggested a technique to identify extremely metal-poor galaxies which share
very similar properties (age, metallicity, star formation rates) with hosts of
LGRBs. The data from \citet{Brown_Kewley_Geller_2008} are plotted in
Fig.~\ref{fig:metallum} as blue circles, together with the observed relations by
\citet{Tremonti_etal_2004}, and by \citet{Richer_McCall_1995} for a sample
of irregular galaxies. Model results for the HOST3 sample are plotted as
magenta asterisks and lie in the same region occupied by the Brown et al. data.
Dark crosses show the corresponding results for the HOST2 sample.  As explained
above, the adopted selection results in clear metallicity cuts in
Fig.~\ref{fig:metallum}. Observational studies of gas-phase metallicity of GRB
hosts could therefore provide important information on the metallicity of the
progenitor stars, although inhomogeneous mixing of metals could complicate the
interpretation. The few objects in the HOST3 samples with high metallicity
are galaxies with very high star formation rate (which coupled with the
instantaneous recycling approximation, results in quite high metallicities of
the inter-stellar medium).

\subsection{The environment of LGRB host galaxies}
\label{sec:environment}

\begin{figure*}
  \includegraphics[scale=0.65,angle=-90]{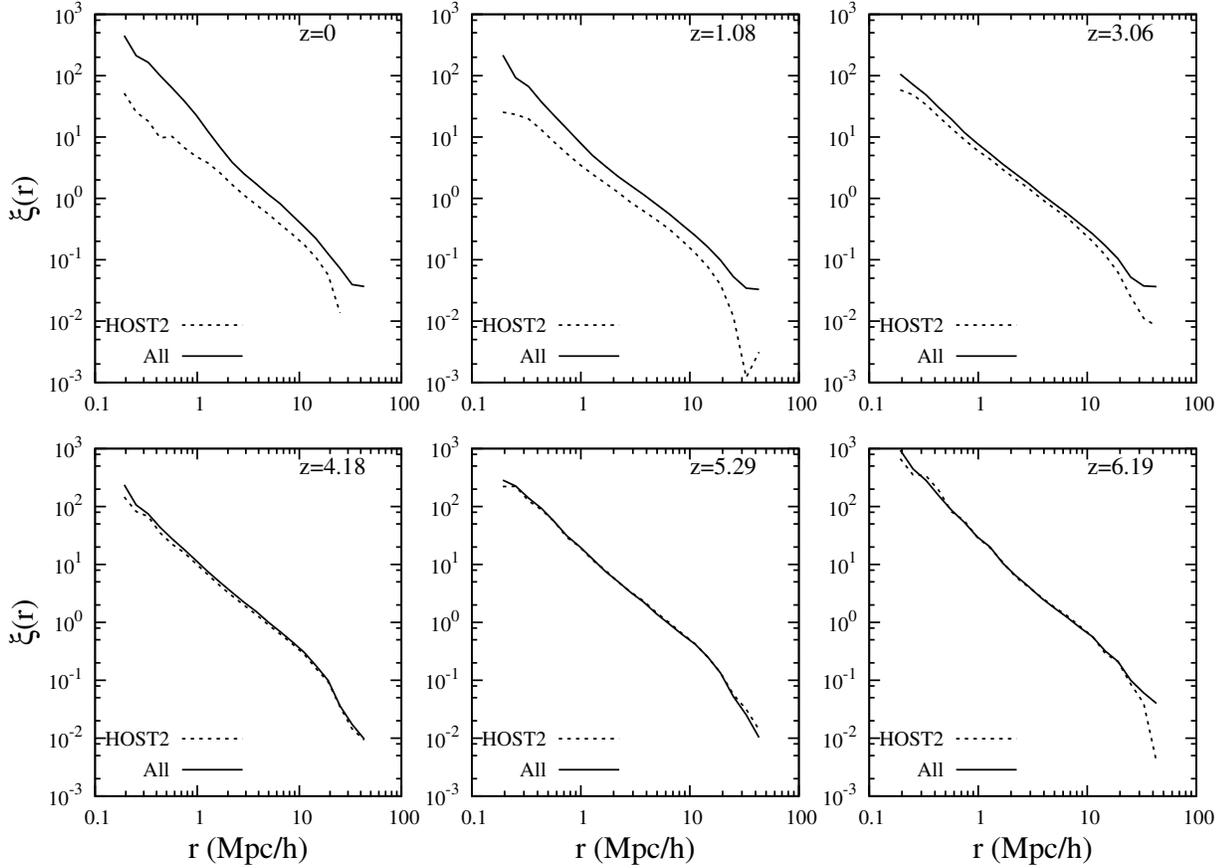}
  \caption {Two-point auto-correlation function for LGRB host galaxies from
  the HOST2 sample (dashed line) and for normal galaxies (solid line), at
  different redshifts, for WMAP3B cosmology.}
\label{fig:cor3}
\end{figure*} 

The physical environment of LGRB host galaxies can provide important
information on the origin of LGRBs. The analysis of GRB hosts environments is, 
however, quite difficult from the observational viewpoint, given the very low
number of identified and well studied host galaxies. Only a few observational
studies have attempted to address this question in the past few
years. \citet{Gorosabel_etal_2003} used photometric redshift information in a
field of $6\arcmin\times6\arcmin$ containing the host galaxy of GRB 000210, and
found no obvious galaxy concentration around the host. \citet{bor04} analysed
the cross-correlation function between host galaxies and surrounding field
galaxies using VLT and public HST data, and concluded that host galaxies do not
reside in high density environments.

The semi-analytic catalogues used in our study provide the positions of the LGRB
host galaxies, as well as of all other galaxies in the simulation box. We have
used this information to study the auto-correlation function of host galaxies,
and the cross-correlation function between hosts and all galaxies in the
simulation (which we will call ``normal'' galaxies). 

In order to compute the two-point correlation function, we adopt the 
\citet{lan93} estimator:

\begin{equation}
  \xi (r) = \dfrac{DD(r)-2DR(r)}{RR(r)}+1\; 
\end{equation} 
where $DD(r)$ is the number of galaxy-galaxy pairs at distance $r$, $RR(r)$ is
the number of random-random pairs, and $DR(r)$ is the number of random-galaxy
pairs. 

Fig.~\ref{fig:cor3} shows the auto-correlation function for LGRB host
galaxies in the HOST2 sample (dashed line) and for normal galaxies (solid
line). The number of galaxies in the HOST3 sample are too low to compute a reliable correlation function. Results are shown for the WMAP3B model, but they are similar for the
WMAP1 simulation. At low redshift, the auto-correlation function of host
galaxies is lower than the corresponding function for normal galaxies. The
difference between the two functions decreases with increasing redshift and the
two functions almost perfectly overlap at $z > 4$. As explained in
Sec.~\ref{sec:sfr}, this is due to the decrease of metallicity at
high redshift which implies that an increasing fraction of the global galaxy
population can host a LGRB event according to our selection (see
Sec.~\ref{sec:method}). 

\begin{figure}
  \includegraphics[scale=0.45,angle=-90]{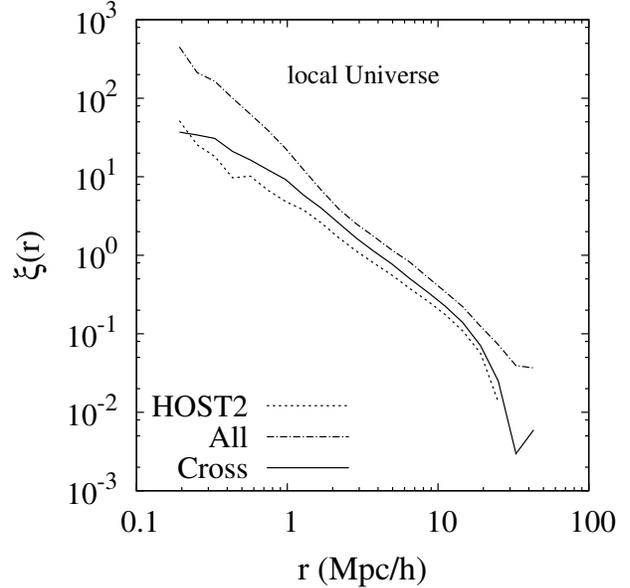}
  \caption {Two-point auto-correlation function for HOST2 galaxies (dashed 
  line) and for normal galaxies (dot-dashed line) in the WMAP3B model. The solid line shows the
  cross-correlation function between host and normal galaxies. } 
\label{fig:cor2}
\end{figure}

The two point auto-correlation functions for the HOST2 sample and for normal
galaxies at $z=0$ is repeated in Fig.~\ref{fig:cor2}, together with the
cross-correlation function between host and normal galaxies (solid line). The
cross-correlation function lies in between the two auto-correlation functions,
suggesting that the probability of finding another host near a GRB host is
lower than the corresponding probability of finding a normal galaxy.
  
Our results are in qualitative agreement with those found by \citet{bor04} and
suggest that LGRB host galaxies tend to populate regions with density lower
than average. This is not entirely surprising if one considers that host
galaxies are typically low-mass star forming galaxies which preferentially live
in low density environments \citep{Kauffmann_etal_2004}.

\subsection{Descendants of high-$z$ LGRB host galaxies}
\label{sec:evolve}

\begin{figure*}
  \includegraphics[scale=0.6,angle=-90]{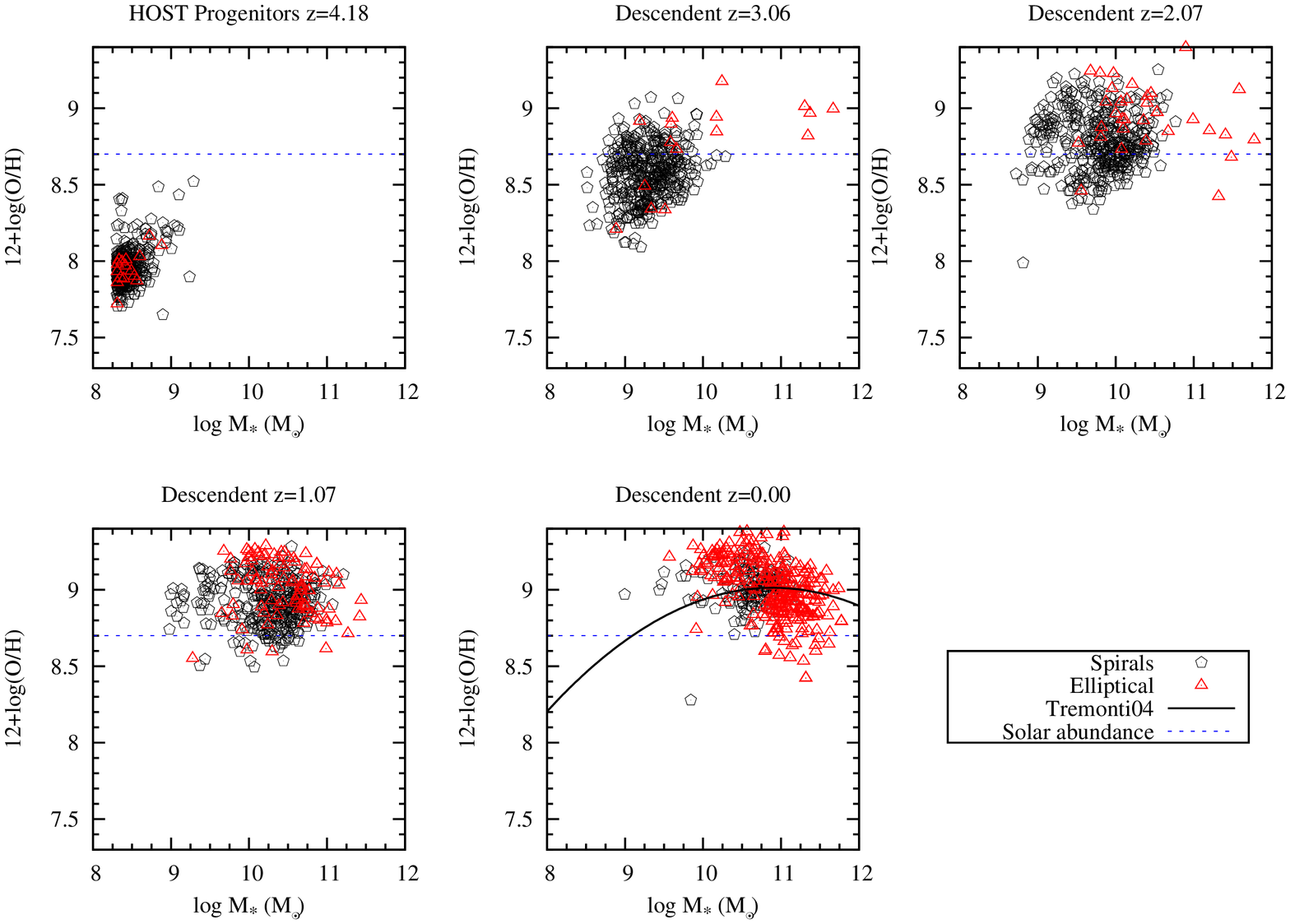}
  \caption{Mass-metallicity relation for HOST3 galaxies at redshift $4.18$
    (top left panel), and for their descendants at lower redshifts in
    the WMAP3B model. The colour
    coding indicates galaxies with different morphologies (red triangles for 
    ellipticals and black diamonds for spirals). The solid black line in the 
    bottom right panel shows the best fit relation found by 
    \citet{Tremonti_etal_2004}. The horizontal dashed line corresponds to the 
    adopted solar abundance.}
\label{fig:c1}
\end{figure*}

From the theoretical point of view, it is interesting to ask which are the
`descendants' of high-redshift LGRB host galaxies. Do they preferentially end
up in massive haloes? What are the typical morphology, colour, mass and
metallicity of the descendants? While this is a very difficult (if not
impossible) question to address observationally, it can be easily addressed 
with the available semi-analytic catalogues, which contain the full merger tree
information for all galaxies in the simulation box.

In this section, we use this information to study the fate of LBRG host galaxies
selected at $z\sim 4$ (hereafter {\it progenitors}) in the observed
mass-metallicity and colour-magnitude planes, and the distribution of host halo
virial mass of the descendant galaxies. For simplicity, we only show results
for our HOST3 sample which contains a lower number of host galaxies than the
HOST2 sample. Results are, however, similar for HOST2.

The top-left panel in Fig.~\ref{fig:c1} shows the mass-metallicity relation for
our HOST3 sample at $z=4.18$. The other panels show the location in the same
plane of the descendant galaxies, down to $z=0$. Galaxies are colour-coded as
functions of their morphological types which are assigned by the ratio of the
total bulge mass to the total stellar mass, $B/T$. Objects with $B/T>0.5$ are
classified as `ellipticals' and are shown as red
triangles, black diamonds represent `spirals' ($B/T<0.5$).

Fig.~\ref{fig:c1} shows that the LGRB host galaxies at redshift $\sim 4$ are
low-mass and low-metallicity galaxies (as discussed in the previous sections),
and that the great majority of them do not have a significant bulge
component. At later time, galaxies grow in mass and their gas-phase metallicity
increases. Fig~\ref{fig:c1} shows that the increase of the gas-phase
metallicity is very rapid: a relatively large fraction of descendants at $z\sim
3$ already have solar metallicity [12+log (O/H) $\sim$8.7, \citealt{all01}] and
most of the descendants at $z\sim 2$ have super-solar metallicity. This
efficient enrichment of the interstellar medium is due to the fact that the
semi-analytic model used in our study assumes a perfect mixing efficiency
of the metals formed by new stars \citep{DeLucia_Kauffmann_White_2004}. The
stellar mass of the descendants evolves more slowly, with very few objects
jumping to masses larger than $10^{11}\,M_{\odot}$, probably as a consequence
of major mergers. At $z=0$, all descendants of LGRBs at $z\sim4$ have
relatively high gas-phase metallicities and cover all the mass range between a
few times $10^9$ and $10^{12}\,M_{\odot}$. Interestingly, a large fraction of
these (about $66$ per cent) have a dominant bulge component.

\begin{figure*}
  \includegraphics[scale=0.56,angle=-90]{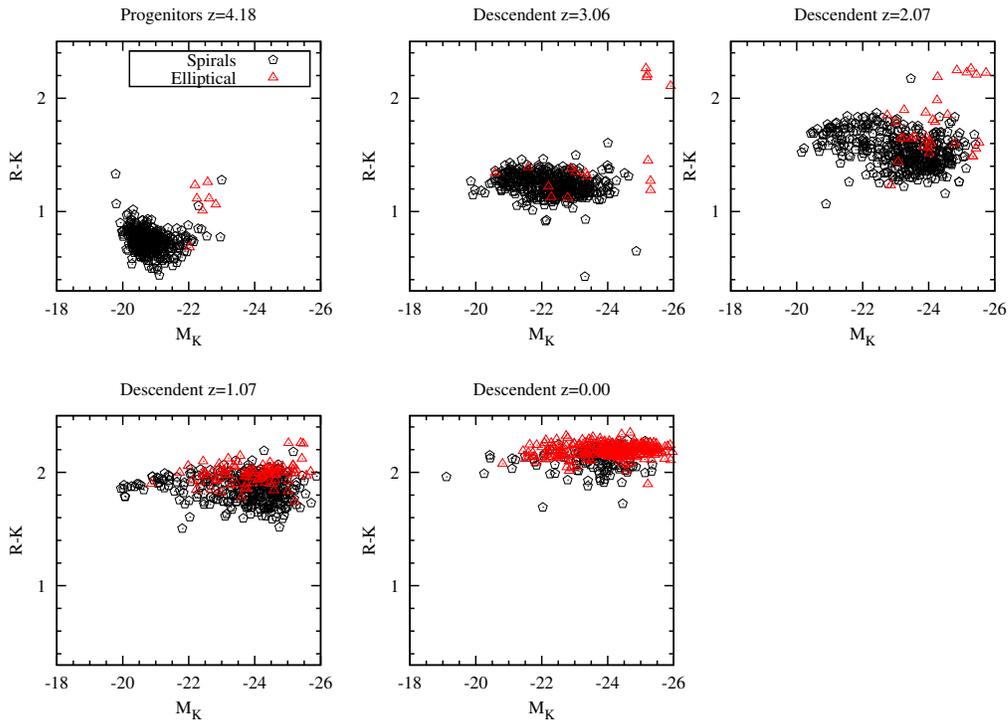}
  \caption {$R-K$ colour as a function of the K-band magnitude for HOST3 galaxies
    galaxies at $z\sim 4$ (top-left panel),  and for their descendants at lower
    redshifts in the WMAP3B model. Red triangles are for elliptical galaxies while black diamonds
    are for spirals.}
\label{fig:c2}
\end{figure*}
Fig.~\ref{fig:c2} shows the location in the colour-magnitude diagram of LGRB
host galaxies at $z=4.18$ (top left panel) and of all their descendants at
later time. LGRB hosts selected at $z\sim 4$ have very blue colours (R-K~$<
1$) and relatively faint magnitudes. The descendants of these galaxies become
progressively redder (they all have R-K~$>2$ at $z=0$) and cover a wider
range of magnitudes at low redshift. A few objects become very luminous and
very red already at $z\sim3$ (these are the same objects which appear very
massive and metal rich at the same redshift in Fig.~\ref{fig:c1}).

\begin{figure*}
  \includegraphics[scale=0.56,angle=-90]{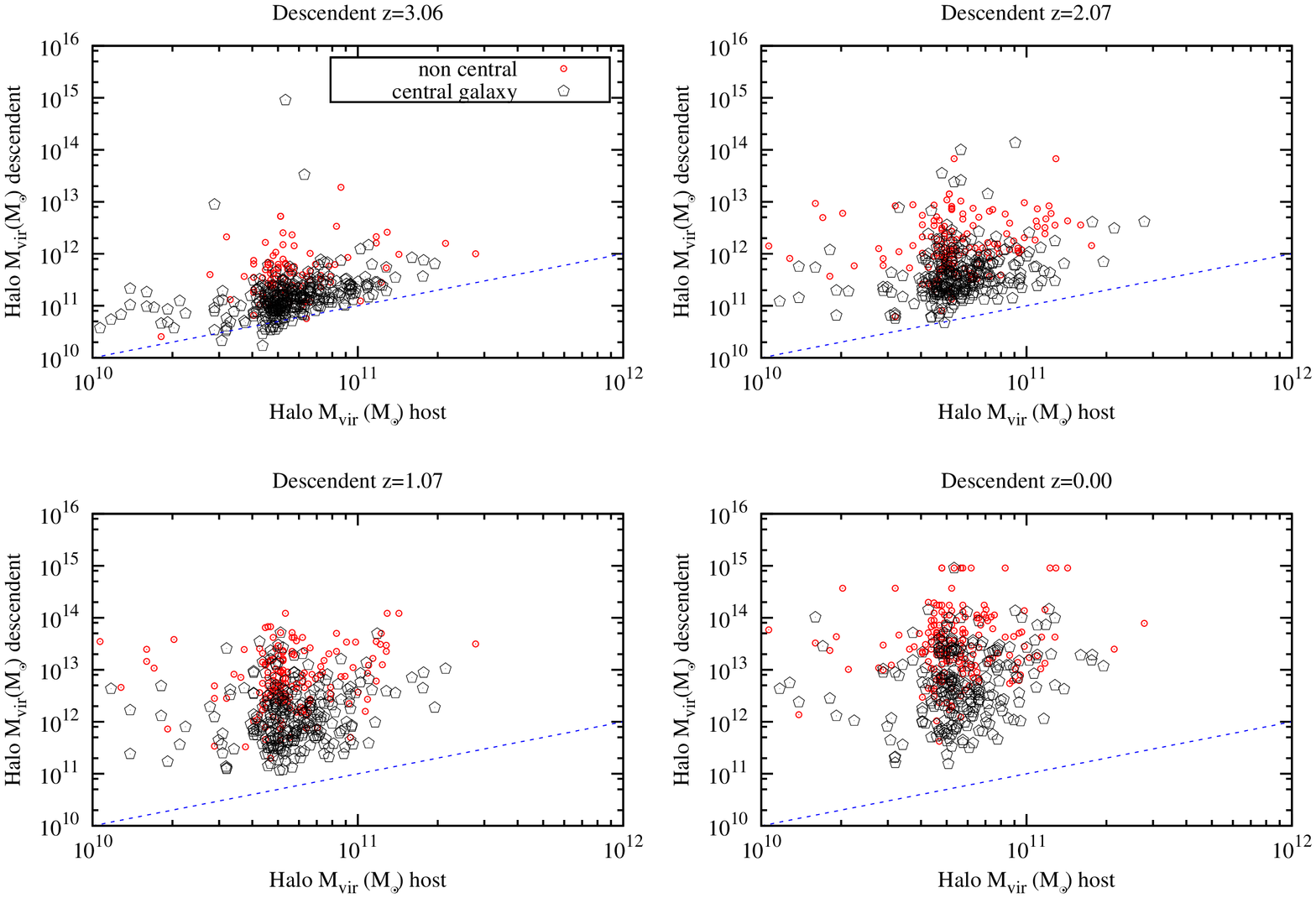}
  \caption{Parent halo mass of all descendants of HOST3 galaxies selected at
    $z=4.18$, as a function of the halo mass of the host galaxies in the
    WMAP3B case. Red symbols are used for satellite galaxies and black symbols for central galaxies.}
\label{fig:c3}
\end{figure*} 
Finally, in Fig.~\ref{fig:c3} we show the parent halo mass of all descendant of
LGRB host galaxies at $z\sim 4$ versus the halo mass of the host. The blue
dashed line in each panel shows the one-to-one relation and is plotted to guide
eyes. As expected, descendant galaxies reside in larger and larger haloes as
time goes on. At $z\sim 2$, the majority of the descendant galaxies still reside
in haloes of mass $\sim 10^{11}\,M_{\odot}$ but a few of them are in relatively
massive haloes ($\sim 10^{14}\, M_{\odot}$) and are satellite galaxies of a
very massive cluster at $z=0$. In the local Universe, the descendants of high-z
LGRB hosts reside in haloes of different mass but most of them still reside in
haloes with mass between $10^{12}$ and $10^{13}\, M_{\odot}$, in agreement with
the results in Sec.~\ref{sec:environment}.


\section{Discussion and Conclusions}
\label{sec:discuss}

In this work, we have studied the physical and environmental properties of
galaxies hosting Long Gamma-Ray Bursts (LGRBs) in the context of a hierarchical
model of galaxy formation. In order to select host galaxies, we have adopted
the collapsar model and used information available from a semi-analytic model
of galaxy formation coupled to high-resolution cosmological simulations (Wang
et al. 2008). 

By imposing different metallicity constraints on the progenitor stars, we have
constructed three host galaxies samples: HOST1 is built without any cut on the
metallicity of progenitor stars of GRBSs, while the HOST2 and HOST3 samples are
constructed by selecting galaxies with progenitor stars of metallicity lower
than $0.3Z_{\odot}$ and $0.1Z_{\odot}$ respectively. 

A number of recent studies have adopted a similar but not identical approach to
study the host galaxies of LGRBs. \cite{Nuza_etal_2007} developed a Monte Carlo
code to identify hosts of LGRBs within a cosmological hydrodynamical
simulations. Their analysis was also based on the collapsar model and limited
to $z \leq 3$. The simulation used in Nuza et al. was relatively small ($10
h^{-1} {\rm Mpc}$) and therefore did not allow a detailed investigation of the
environmental properties of LGRB hosts. As confirmed by our study, they
pointed out that if LGRBs are generated by the death of massive young stars of
metallicity lower than a certain value, they do not provide a good tracer of
the cosmic star formation history (see Sec.~\ref{sec:sfr} and later). 

\cite{Lapi_etal_2008} used the cosmological star formation rate below a
critical metallicity to estimate the event rate of LGRBs. To this purpose, they
employed the galaxy formation model presented in Granato et al. (2004). Lapi et
al. find that their predicted number counts of LGRBs agrees well with the
bright {\it SWIFT} data, without the need for an intrinsic luminosity
evolution. They find that host galaxies are dominated by young stellar
populations, are gas rich and metal-poor. The model adopted in Lapi et
al. (2008) does not follow galactic disks nor does it consider mergers between
galaxies or between haloes. In addition, the model does not provide spatial
information of model galaxies.

In another recent study, \cite{cou07} used N-body/Eulerian hydrodynamic
simulations, and identified GRB host galaxies with those having SFR and
SFR-to-luminosity similar to those of 10 observed GRB hosts in the redshift
range $0.43<z<2.03$. They found that the host galaxies have low stellar masses
and low mass-to-light ratios, are young and bluer than typical galaxies at the
same cosmic epochs. Their identification of simulated galaxies with observed
hosts is limited by uncertainties on the observational estimates of the SFR and
luminosity of host galaxies. In addition, their simulated volume is relatively
small ($32\,h^{-1}\,{\rm Mpc}$) and the analysis is limited to relatively low
redshifts. 

Compared to previous works, our study uses a much larger simulated volume (a
box of $125\,{h}^{-1}\,{\rm Mpc}$ on a side - see Sec.~\ref{sec:sam}) that
allows us to study the environmental properties of model host galaxies. In
addition, the use of two different cosmological models allows us to analyse the
dependence of results on cosmology. The semi-analytic model employed in our
work has been studied in a number of previous papers, and it has been shown to
successfully reproduce a number of observational results for the global galaxy
population in the local Universe and at higher redshift (see
Sec.~\ref{sec:sam}).

In agreement with previous work, and as expected due to the global increase of
the ISM metallicity with decreasing redshift, we find that when assuming a
metallicity threshold for progenitor stars of LGRBs, they do not represent a
perfect tracer of the cosmic star formation history. The bias is stronger as the
metallicity threshold assumed is lowered. At higher redshift, the cold-phase
metallicity of model galaxies is lower, and galaxies form stars at higher
rates. As a consequence, there is a higher rate of LGRBs per galaxy and the
host galaxy population includes a larger fraction of the global galaxy
population, with the two populations sharing very similar physical properties.

At lower redshift, the host galaxy population is dominated by galaxies with
low-masses, relatively young ages, blue colours, and luminosity below $L_*$, in
qualitative agreement with observational measurements. We note, however, that
while $\sim 90$ per cent of the galaxies in our simulations have stellar masses
lower than $\sim 10^{10}\,{\rm M}_{\odot}$, only $\sim 0.3$ per cent of them
are in the HOST3 sample and  $\sim 13$ per cent in the HOST2 sample.

The metal content and metallicity evolution of host galaxies also appear to be
in agreement with observational estimates. Since in the model adopted in this
study stars form with the metallicity of the cold-phase component at the time
of the star formation, host galaxies have average metallicity lower or close to the
metallicity threshold adopted. Metallicity studies of LGRB host galaxies could
therefore provide important constraints on the adopted collapsar model. It is
important to note, however, that inhomogeneous mixing of metals (which is not
considered in the model adopted here) could significantly complicate this
picture. The metal content of host galaxies does not significantly evolve with
redshift, while the overall galaxy population exhibit some weak evolution.

Taking advantage of the spatial information provided by the semi-analytic model
used in our work, we study the clustering properties of model host galaxies and
compare them to the corresponding properties of the global galaxy population. 
Since a larger fraction of the whole galaxy population can host LGRBs at high
redshift, the clustering properties of host galaxies do not differ
significantly from the clustering properties of `normal' galaxies. At lower
redshift, the host galaxies are significantly less clustered than normal
galaxies, as expected due to their physical properties. Interestingly, we find
that the cross-correlation function between host and normal galaxies lies in
between the the auto-correlation functions of host galaxies and normal
galaxies, suggesting that the probability of finding another host galaxy nearby
a GRB host is lower than the corresponding probability of finding a normal
galaxy. This implies that LGRBs reside in regions with density lower than
average. When larger samples of LGRB host galaxies will become available, it
will be possible to test our clustering predictions.

Using the available semi-analytic catalogues, which contain the full merger
tree information for all galaxies in the simulation box, we find that LGRB
host galaxies at redshift $\sim 4$ (which have low stellar mass,
low metallicity and no significant bulge component) evolve into relatively
massive, red galaxies at redshift zero. While the descendants of high-z LGRBs
reside in haloes of different mass in the local Universe, most of them still
sit in haloes with mass between $10^{12}$ and $10^{13}\,{\rm M}_{\odot}$.

\section*{Acknowledgements}
We are indebted to Dr. Jie Wang for making available their simulated galaxy
catalogues and simulation outputs. We thank Sandra Savaglio for useful discussions. 
SM would like to acknowledge the Humboldt Foundation for travel support.
We also acknowledge an anonymous referee for a constructive report that improved the paper.

\bibliographystyle{mn2e}
\bibliography{paper}

\end{document}